\documentclass{article}
\usepackage{graphics}
\usepackage[dvips]{graphicx}
\usepackage{bm}
\usepackage{latexsym}
\usepackage{pxfonts}
\begin{document}

\title{Length scale of dynamic heterogeneity in polymer nanocomposites.}
\author{S. Srivastava, A. K. Kandar, J. K. Basu, M.K. Mukhopadhyay, L. B. Lurio, S. Narayanan and S.K. Sinha.\\
\emph{\small{Department of Physics, Indian Institute of Science, Bangalore, 560012, India}}\\
\emph{\small{Department of Physics, University of California San Diego, La Jolla,  CA 92093, USA,}}\\
 \emph{\small{Department of Physics, Northern Illinois University, De Kalb, IL 60115 USA,}}\\
\emph{\small{Advanced Photon Source, Argonne National Laboratory, Argonne, IL 60439, USA.}}}

\maketitle
\abstract{\emph{
Using X- ray photon correlation spectroscopy measurements on gold
nanoparticles embedded in polymethylmethacrylate we provide evidence
for existence of an intrinsic length scale for dynamic heterogeneity
in polymer nanocomposites similar to that in other soft materials.We also show how the dynamics varies in a complex way with various parameters.}}\\%

Polymer nanocomposites (PNC) are a novel class of multifunctional
hybrid materials which are obtained by appropriate mixing of
nanoparticles and polymers leading to a wide range of applications
in terms of their unique electrical, optical, and thermo-mechanical
properties [{ 1 - 4 }]. They also belong to this new class of
materials, under the umbrella of soft glassy materials, which
exhibit rich and complex thermal, mechanical and rheological
behavior [{ 5 - 11 }]. Extensive experimental work has been performed
over the decade or so on studying the modifications of thermal and
mechanical properties imparted to the matrix polymer by addition of
nanoparticles of various shapes and sizes [{4,12-15}]. Numerical
simulations and theoretical calculations have started to emerge in
the meanwhile to provide some insight and understanding into the
static structure and phase behavior in various model nanoparticle -
polymer hybrid systems [{16-20}]. Homogeneous dispersion of
nanoparticles in polymer matrices is a major obstacle [{3}] towards
the ultimate goal of obtaining high performance materials. In this
respect it has been observed by us [{4}] and a few other groups
[{21}] that using a polymer capped nanoparticle in a matrix of
chemically matched homopolymer matrix is a very effective way of
ensuring homogeneous dispersion. However, the theory and simulations
for the static structure and phase behavior for such systems are
only just beginning to emerge [{16-20}] and a microscopic theory to
treat the dynamics, of both polymer capped and uncapped
nanoparticles in polymer matrices, including viscoelasticity, glass
formation or diffusion of nanoparticles does not exist. Dynamics of
soft glassy materials is extremely rich and complex [{5,7,22-24 }].
Depending on various parameters like volume fraction, ($\phi$) and
temperature , ($T$) such systems exhibit complex slow dynamics
characteristic , of glassy or jammed phase [{5,7, 22-24 }]. Dynamic
heterogeneity is a key feature of such systems and evidence of
length scale dependent dynamic heterogeneity has been observed in
simulations  [{ 10,25-26 }]and certain experiments [{27, 28}]. In this communication we provide evidence for existence of a length scale of dynamic heterogeneity for PNC. We also provide the first glimpse of the complexity and richness of phase behavior in dynamics of polymer nanocomposites as a function  of  $\phi$, T and the wave vector, q , through detailed synchrotron multi-speckle x-ray photon correlation spectroscopy (XPCS)
measurements. \\
In XPCS one measures the intensity autocorrelation
function [{9,11,14,29}],

{\begin{equation}
g_2(q,t)= 1 + b{\vert f(q,t)\vert}^2.
\end{equation}}\
Here, $f(q,t)$ is the intermediate scattering function (ISF), $ b $
is the speckle contrast and t is the delay time. We have used ISF of general form,
{\begin{equation}
f(q,t) = exp-{(t/\tau)}^\beta,
\end{equation}}\
where $\tau$ and $\beta$ represents the characteristic relaxation
time of the system and the Kohlrausch exponent respectively.
The mean relaxation time $\tau$ follows a power law behavior and is given by
{\begin{equation}
\tau \propto q^{-\alpha}.
\end{equation}}
For diffusive motion, $\alpha$ = 2. However, for ballistic or
super-diffusive motion  $\alpha$ is predicted to be $\sim$ 1 and for
sub diffusive motion $\alpha > 2 $ [{30}].
 Normally , $\beta$ is
independent of $q$, however there has been some recent observation
in  colloidal gel systems where $\beta$ has also been found to
depend on $q$. However the underlying physics is not clear. We performed XPCS measurements at 8-ID beamline of the Advance Photon Source with  partially coherent X-rays of energy 7.35 KeV of beam size
20$\mu$m X 20$\mu$m using a CCD (Princeton Instruments). The total exposure time at each sample position was limited to $\sim$ 10 minutes to minimize radiation damage.
 The results of XPCS measurements, presented here, are based on three different samples (A,B,C), of gold nanoparticles capped with
 Polymethylmethacrylate (PMMA, 120K , sigma aldrich) and embedded in PMMA matrix, prepared by a method described earlier [4].
All samples were annealed at ${ 150^oC}$ for 24--30 hrs under a vacuum of 5 X 10$^{-4}$ mbar.
\begin{figure}
\center{\includegraphics[scale=1.1]{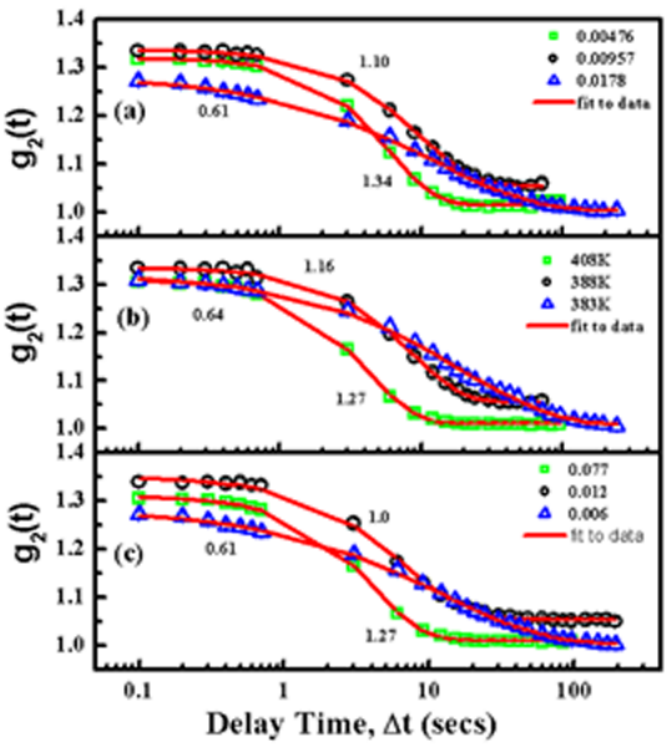}
\caption {\label{fig:epsart3} $g_2$(t) vs t  showing cross over in $\beta$  values with respect to (a) Wave vector, q (b) Temperature, T and (c) Volume fraction, $\phi$ for samples A , B and C. The relaxations not only show simple exponential ($\circ$ ) decay but compressed ($\square$) and stretched ($\bigtriangleup$) exponential  decay as well. Solid line are fits with  eqn. 1 . Number indicated side by are the $\beta$ values extracted from the fits.}}
\end{figure}
 Sample A consists of gold nanoparticles of mean diameter, d, 5 nm embedded in PMMA matrix  with
 volume fraction ($\phi=$ 0.006 ) corresponding to a mean interparticle spacing , h $\sim$ 196 $\AA$.
 The glass transition temperature ,  (T$_g$), for this sample as measured using modulated differential
 scanning calorimetry, was found to be 392 K . Similarly for sample B, d is 6 nm, h $\sim$ 180 $\AA$,
 $\phi=$ 0.012 and T$_g$$\sim$ 380 K. Sample , C is embedded with nanoparticle of diameter 10 nm ,
 with h $\sim$ 105 $\AA$, $\phi=$ 0.077 and T$_g$ $\sim$ 381 K.
 Fig.1 shows the rich and complex dependence of dynamics in such
 nanocomposites on various parameter like volume fraction,
 temperature and crucially the wave vector of measurement and hence
 the length scale. Since x-rays are mostly sensitive to the
 scattering contrast between gold nanoparticle and PMMA , the
 observed dynamics is mostly a reflection of the motion of the
 PMMA capped gold nanoparticles through the background of matrix
 PMMA and is essentially sensitive to the length scale
 dependent viscoelastic property of the medium.  It is clear that the relaxations
 evolve  as a fraction of the various parameters  $\phi$ , T and q
 for all the samples and shows stretched - compressed cross over as
 has been observed for some other soft matter samples [11,25].

\begin{figure}
\center{\includegraphics[scale=1.1]{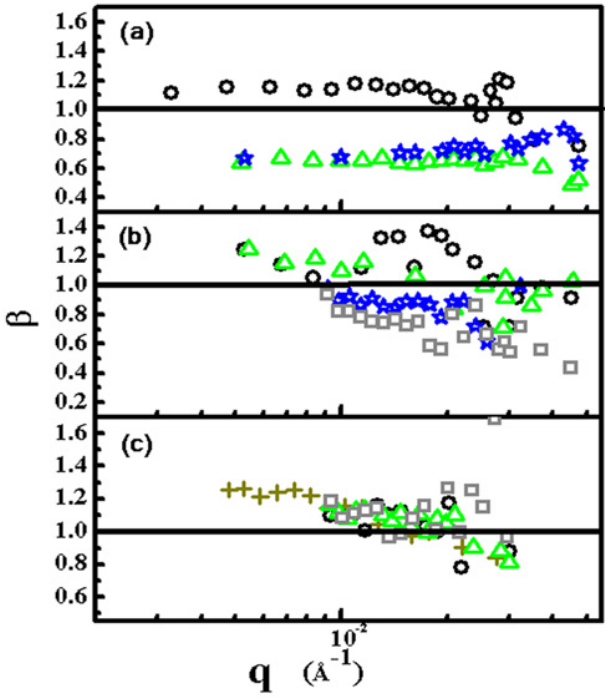}
\caption{\label{fig:epsart1} $\beta$ vs $q$ at various temperatures
for  samples, A, B, C shown in panel a, b, c respectively . The
respective temperatures are ($+$ - 408K,  $\circ$ - 398K,
$\bigtriangleup$ - 388K, $\bigstar$ - 383K,  $\square$ - 378K).}}
 \end{figure}

 However much
 better insight about complexity of the dynamics is obtained by looking at the dependence of the
 $\beta$ and $\tau$, extracted using eqn .2  on these parameters.  In Fig. 2 we show how the  obtained exponent ,
 $\beta$ ,for various samples at different respective measured temperatures vary as a function of wave vector $q$.
Let us first  discuss the low q  behavior of $\beta$ observed in our
samples. Clearly , for sample A , the relaxation is highly stretched
, for measured temperatures, $T$,  below the thermodynamical glass
transition temperature , T$_g$, as expected. However for $T$ $>$
$T_g$ we find the ubiquitous $\beta$ $>$ 1 behavior indicative of
anomalous diffusion typical of slow relaxation in soft and granular
materials undergoing jamming transition [{5, 7, 10-11, 22-24,
27-28}].

\begin{figure}
\center{\includegraphics[scale=1.2]{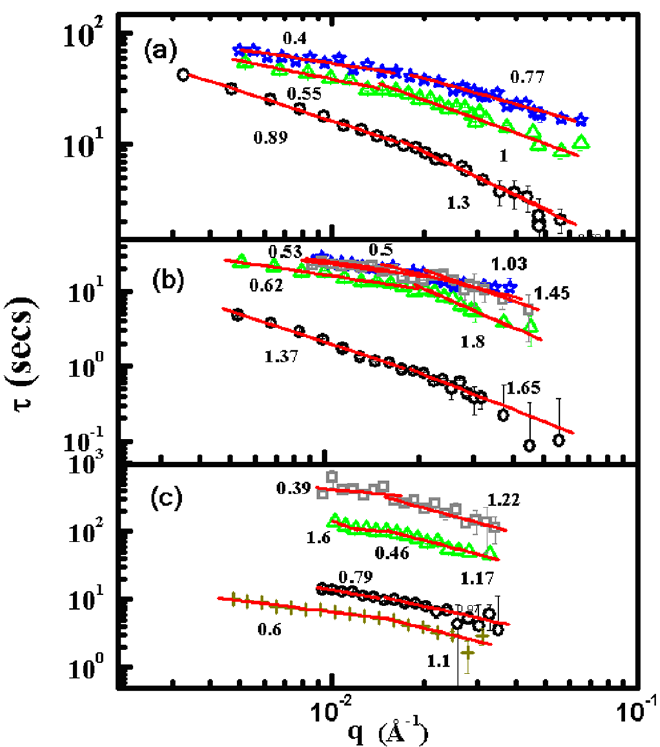}
\caption{\label{fig:epsart2} Relaxation time , $\tau$, vs $q$ for
samples A, B, C  in panel a, b, c respectively, at different
temperatures ($+$ - 408K,  $\circ$ - 398K, $\bigtriangleup$ - 388K,
$\bigstar$ - 383K,  $\square$ - 378K).The straight lines are
linear fits to the data used to estimate $\alpha$, indicated along side each curve according to eqn. 3}}
\end{figure}
 On the other hand, for sample C, $\beta$ shows a clear wave vector dependent behavior. At low q,  $\beta$ $ > $ 1 for all measured temperatures ,
 including that below its , $T_g$ , indicating that the dynamics in this system is
  dominated by viscoelasticty of the effective medium due to the surrounding particles and polymers . This behavior has remarkably similarity to the predicted behavior by Bouchaud et al [23].
  Interestingly sample B, which has intermediate volume fraction clearly shows evolution in terms of
   the relaxation spectra as a function of measured temperature .
Specifically , at low q , (q $\leq$ 0.02$\AA$$^{-1}$) $\beta$  $\leq$  1 for T $\leq$ $T_g$ while $\beta$ $>$ 1 for T $>$ $T_g$. The value of $\beta$ are closer to 1 for T $\leq$ $T_g$ for sample B than it is for sample A and also increases with temperature. Thus the dynamics in such systems is intermediate between that of sample A and B and the intimate relation of T and $\phi$ in determining the dynamics of PNC is vividly demonstrated
here. At higher $q$ (q $\geq $  0.02 $\AA$$^{-1}$,  $\beta$ $\sim$ 1
for all T and $\phi$ , except for the case of sample A at T $<$
$T_g$ . The  cross over from $\beta$$\sim$1 to $\beta$$>$ and $<$ 1
takes place at wave vector , $q_c$. The temperature dependent cross
- over in $\beta$ is opposite to what has been observed recently [{
11}]  and indicates fundamentally different dynamics as compared to
nanoparticles in molecular supercooled fluids. The cross over in
$\beta$ for $q$ $<$ $q_c$ as a function of temperature (for sample A
and B) can be understood as a cross - over from polymer dominated
dynamics at low temperatures (below $T_g$ of PMMA = 388 K) to
particle dominated dynamics at T $>$ $T_g$. However deeper insight
can be obtained from Fig. 3 which quantifies the wave vector
dependence of relaxation time $\tau$ , and delineates the nature of
microscopic diffusion in our PNC system. We find that similar to $\beta$, $\alpha$ also shows a cross over length scale , corresponding to $q_c $ $\sim$ 0.02 $\AA$ $^{-1}$. For $q>$ $q_c$ we find that $1$ $\leq$ $\alpha$ $\leq$ $2$ depending on T and $\phi$ of nanoparticles, whereas for $q$$<$ $q_c$ , 0 $<$ $\alpha$ $<$ 1. This is shown in Fig. 4(a) for all our samples and at all measured temperatures . This low wave vector behavior,
coupled with non - exponential relaxation indicates onset of a form
of dynamic heterogeneity in the PNC systems.
If we define $l_c$ $\propto$ 1/$q_c$ as a length scale corresponding to the cross - over wave vector then this provides an estimate of the smallest
region in the system where dynamics is spatially heterogeneous. For
sample C we observe that this length scale increases slightly ($q_c
\sim$ 0.01$\AA$ $^{-1}$). Our measurements extends only upto $q\sim$
0.005 $\AA$$^{-1}$ and hence $l_{max}$ $\propto$ 1/$q_{min}$
provides an estimate of the largest size of such domains. It might
be noted here that qualitatively similar behavior of length scale
dependent relaxation was also observed in recent molecular dynamic
simulation of colloidal gels and a similar length scale was
identified at where there is a cross over in the dynamics[{25}].
This implies that the exponent $\alpha$ is not a constant over the
entire q range.This fact is highlighted in Fig. 4(b) where we have
identified various regions in q space and the corresponding values
$\alpha$ could take which includes regions within our measured range
($q_{min}$ $<$ q $<$ $q_{max}$ ) and outside (refer caption).
The existence of the hydrodynamic region with $\alpha$ $=2$ can be
conjectured to exist at wave vectors smaller than our $q_{min}$. It
might  be interesting to note that for one of our data ( sample C, T
= 388K),  we clearly see three regions for $\alpha$ including a
region at low q  where $\alpha$ $\sim$ 1.6 indicative of the above
mentioned onset of  hydrodynamic region. It has been observed that
[{31-32}] an  intrinsic length scale for a supercooled liquid
corresponding to onset of dynamic heterogeneity $l_{CRR}$ exists
below which diffusion is anomalous and $\alpha$ $\rightarrow$ 0,
while above it normal diffusion with $\alpha$ $\rightarrow$ 2
exists. Depending on the temperature and especially near and below
$T_g$ there is an intermediate region where $\alpha$ takes on values
between 0 and 2 [{25, 31 - 32}].
\begin{figure}
\center{\includegraphics[scale=1.1]{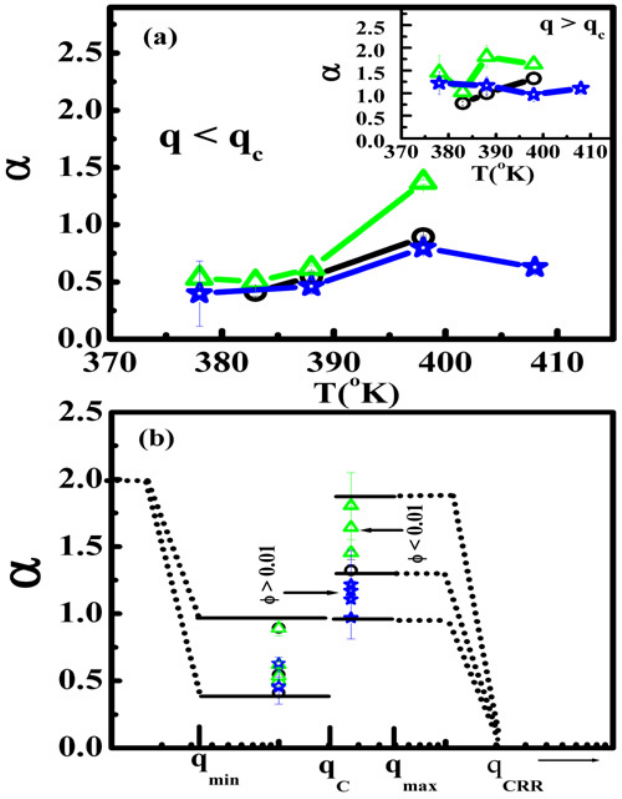}
\caption{\label{fig:epsart3}(a): $\alpha$  vs T , obtained from
linear fits in  fig. 3 for different samples A ($\circ$ ), B
($\bigtriangleup$), C ($\star$), for q $<$ $q_c$. Inset shows
$\alpha$ vs $T$ for q $>$ $q_c$ for three samples, A, B and C (
symbols defined above). (b): Schematic of our model for variation of
$\alpha$ within certain wave vector ranges as defined in text.
$q_{max}$  is the highest measured q in our data . The solid lines
defines the band of $\alpha$ values within our measured q range
$q_{min}$ $<$ q $<$ $q_{max}$ while dotted lines represent expected
behavior outside this range. $\alpha$ lies mostly within 1 in the
band  $q_{min}$ $<$ q $<$ $q _c$ and 1 $\leq$ $\alpha$ $<$ 2 for
$q_c$ $<$ q $<$ $q _{max}$. The dotted line beyond $q_{max}$ indicates projected variations of $\alpha$ with q ( $\alpha$ $\rightarrow$ 0, q $\rightarrow$ $q_{CRR}$ ) for supercooled liquids (ref 31, 32) near and below $T_g$.}}
\end{figure}
 It is reasonable to assume that
$l_{CRR}$ $\sim$ $\xi$$_{CRR}$, which for our PNC has been measured
to be $\sim$ $25$ $\AA$ [4]. This corresponds to a $q_{CRR}$ of
$\sim$ 0.3$\AA$$^{-1}$ which is not reachable in a typical XPCS
measurements. In our case $q_{CRR} \gg q_{max}$ (refer caption).
Hence the values of $\alpha$$\rightarrow$2 for $q$ $>$ $q_c$
corresponds to this region ( $q_c$ $<$ $q$  $<$ $q_{CRR}$ ) for the
polymer glass. For our PNC system $\alpha$ $<$ 2 for q $>$ $q_c$
since particle induced jamming is prevalent. For the highest $\phi$,
we have used for our measurements, this jamming effect increases to
such an extent that the cross over in dynamics is not clearly
observed within the measurable $q$ range (Fig. 3(c)and Fig. 4(b)).
Clearly our systems not only exhibit some of the conventional phases
found in other related soft and glassy systems but exhibits new
phases in dynamics which, to our knowledge, has not been found
earlier. The above discussion can be summarised by identifying different regions in the T - $\phi$ - q  parameter space in which our PNC shows distinct dynamical behavior. In Phase $I$ the dynamics is typical of that observed in jammed systems and in soft glassy matter, in general, and is
characterised by a compressed exponential relaxation and hyper -
diffusive  motion ( $\alpha \sim$ 1) . This type of dynamics is seen mostly at $T > $ $T_g$ and  $ q < q_c$ for sample A and B , which
have relatively low volume fraction of gold nanoparticles. In Phase
$II$ the dynamics is unusual  since $\beta > 1 $ and $\alpha  < $1 . This phase is only seen for sample C, with the highest volume fraction of our samples,  at all temperatures and indicates that the dynamics is dominated by particles induced jammed behavior. The diffusion is highly anomalous with weak wave vector  dependence of relaxation indicating heterogeneous
dynamics  mainly due to particle jamming. Phase $III$ with $\beta
\sim$ 1 and $\alpha > $ 1  is seen for $q > $ $q_c$ and $T > T_g$
and is indicative of liquid like behavior, as expected. Phase $IV$
( $\beta < $ 1 and $\alpha > $1 ) occurs for sample A and B for $ q
> q_c$ and T $ < T_g$ . As explained  earlier, this is the phase
which would have been the hydrodynamic region ( $\tau \propto$
1/$q^2$ for q $ < $  $q_ {CRR}$). However the onset of the
additional dynamic heterogeneity in PNC truncates this region to
within  $ q_c < q < q_{CRR}$ . It is possible that the measurements at temperatures T $>>$ T$_g$ would ultimately reveal the 1/q$^2$ diffusive dynamics as revealed in various other measurements [11,25]. Finally, in Phase V,  $\beta$ $<$ 1
and $\alpha$ $<$  1 which occurs for sample A and B for T $<$ $T_g$
and $ q < q_c $. The dynamics in this phase is dominated mostly by
polymer visco-elasticity but the heterogeneity in dynamics is
intrinsic to the PNC system due to the creation of domain of size
$l_{min}$ $ < l < l_c$. We have presented systematic studies of dynamics in typical PNC system consisting of PMMA capped gold nanoparticles embedded in PMMA matrix of identical molecular weight with various volume
fractions. Our temperature and wave vector dependent XPCS measurements provides new insight into the fascinating dynamics in these systems and for the first time reveals the existence of a dynamical heterogeneity length scale intrinsic to PNC systems. The phase behavior in such systems is more complex and interesting and requires extensive investigation  to fully understand the physics at the microscopic scale leading to such
diverse and puzzling behavior.\\
The authors acknowledge M. Sprung (APS) for discussions
and A. Sandy (APS) for assistance in experiments.This work
benefitted by the use of facilities  at APS, which is supported by
U.S. DOE (BES) under Contract No. W - 31 - 109 - Eng - 38  to the
University of Chicago. Part of the work has been supported by DST, India and UCSD.

\end{document}